\documentclass[preprintnumbers,amsmath,amssymb,showpacs,twocolumn]{revtex4}

\usepackage{graphicx}
\usepackage{dcolumn}
\usepackage{bm}

\begin{document}

\title{Probabilistic Teleportation of Two-Particle State of General Formation in Ion Trap}

\author{LIAN Shi-Man, YAN Feng-Li}\thanks{To whom correspondence should be addressed. Email: flyan@mail.hebtu.edu.cn\\Supported by the National Natural Science Foundation of China under
Grant No: 10971247, Hebei Natural Science Foundation of China under
Grant No: F2009000311.}

\affiliation { College of Physics Science and Information
Engineering, Hebei Normal University, Shijiazhuang 050016,
China\\Hebei Advanced Thin Films Laboratory,  Shijiazhuang 050016,
China}

\date{December 28, 2009}

\begin{abstract}
We propose a scheme for probabilistic teleportation of an
unknown two-particle state of general formation in ion trap. It is shown that one can
realize  experimentally this teleportation protocol  of
 two-particle state with presently
available techniques.
\end{abstract}


\maketitle

Entanglement is the most peculiar feature of quantum physics  and
lies at the heart of quantum information \cite{NielsenChuang}. One
of the most striking application of entanglement is quantum
teleportation \cite{s1}, which says that  an unknown state can be
transmitted
 from a sender Alice  to  a spatially distant receiver Bob with the aid of classical communication and
  a  previously shared entanglement.  Quantum teleportation  has been widely used in the development of quantum
computation and quantum communication
\cite{s1,s2,s3,s4,s5,s6,BDEFMS, DINGShengChaoJINZhi,zhangguohuayanfengli}.

 In the original protocol of Bennett et al. \cite {s1}, the sender Alice and a spatially distant receiver Bob
 are sharing  an Einstein-Podolsky-Rosen (EPR)
 pair first. Then  Alice makes  the Bell state measurement and transmits her measurement outcome to Bob via a classical
 channel. According to Alice's measurement result, Bob performs a
 corresponding unitary transformation. After that the quantum state
 has been teleported successfully. Here Alice
  does not know either the state to
 be teleported or the location of the
  intended receiver, Bob. Bennett et al \cite {s1} also gave a protocol involving teleporting
  an unknown state of a qudit via a maximally entangled state in $d\times d$
  dimensional Hilbert space and by transmitting $2{\rm log}_2d$ bits of
  classical information.

 Since the seminal work of Bennett et al. \cite {s1}, a lot of work has been done in the field of quantum teleportation \cite {Braunstein, VaidmanPRA1994, BKprl1998,
  GRpra,  KB, s7, GYLScienceinChina2, WangYan,APpla, PAjob, DLLZWpra2005,Zhangzhanjun,TIANDongPingTAOYingJuanQINMeng,
ZHANGXinHuaYANGZhiYongXUPeiPei, LiXHandDengFG,ZHANZHANGQunWANGMA,ZHANGLIUZUOZHANGZHANG,QIANFANG,ZHOUYANGLUCAO,DONGTENG} and quantum
teleportation has been experimentally demonstrated by several groups
\cite {BPMEWZ,FSBFKP,NKL,Boschi}. Furthermore, quantum teleportation
 has also been generalized to  more general situations,
  for example, where two parties may  share not with a set of pure entangled states,
  but with a noisy quantum channel.
 In this situation they  can  use an error correcting code \cite {Gottesman}, or  they
  can  share the entanglement through  this noisy channel  and then  use teleportation \cite {BBP96}.

  It has been demonstrated that for infinite dimensional
Hilbert spaces, quantum teleportation is also possible \cite
{Braunstein,
  VaidmanPRA1994, BKprl1998}.

 Karlsson and Bourennane proposed the so-called
 controlled quantum
 teleportation protocol \cite{KB,DLLZWpra2005,Zhangzhanjun,s7,GYLScienceinChina2,LiXHandDengFG}.
 In  the protocol, one can perfectly transport an unknown state
  from one place to another  via a previously shared
Greenberger-Horne-Zeilinger (GHZ) state using local operations and
classical communications  under the control  of  a third party. The
unknown quantum  state can not be teleported  unless the third party
permits them to transmit the state.

In the situation where  the shared entanglement between the sender
Alice and the receiver Bob is not in a maximally entangled state
then they cannot teleport a qubit with both unit fidelity and unit
probability. However, it  is possible to have unit fidelity
teleportation but with a probability less than unity by using a
non-maximally entangled state.  This is so called probabilistic
quantum teleportation \cite {APpla,PAjob}, and has been shown to be
possible using a non-maximally entangled basis as a measurement
basis. Furthermore, this probabilistic scheme has been generalized
in two ways: (i) to enable teleportation of $N$ qubits \cite {GRpra}
and (ii) controlled teleportation \cite
{s7,GYLScienceinChina2,LiXHandDengFG}.

So far, many schemes have been proposed about the teleportation of
two-particle state \cite{ShiJiangGuo,LuGuo,DaiChenLi,yantan}. But
most of them settled at mathematic logic level and did not present a
concretely physics system. Solano et al. \cite
{SolanoCesarMatosFilho} propose a method for implementing a reliable
teleportation protocol of an arbitrary internal state in trapped
ions, and Home et al. \cite{HomeStean} proposed a scheme for
implementing optimal probabilistic teleportation between two
separated trapped ions. In this Letter we propose a scheme for
probabilistically teleporting an unknown two-particle state in ion
trap.

We suppose that ion 1 and ion 2 in trap $A$ are in  an arbitrary
electronic state
\begin{equation}
|\phi\rangle_{12}=\alpha
|ee\rangle_{12}+\beta|eg\rangle_{12}+\gamma|ge\rangle_{12}+\delta|gg\rangle_{12},
\end{equation}
where $|\alpha|^2+|\beta|^2+|\gamma|^2+|\delta|^2=1$; $|g\rangle$ and $|e\rangle$ denote ground state and excited  state of the ion, respectively.  The ions 3, 4,
5 and 6 are in trap $B$.  The ions 3 and 4 is in the following state
\begin{equation}
|\psi\rangle_{34}=a|ee\rangle_{34}+b|gg\rangle_{34},
\end{equation}
and the state of ions 5 and 6 is
\begin{equation}
|\psi\rangle_{56}=c|ee\rangle_{56}+d|gg\rangle_{56}.
\end{equation}
Here $|a|^2+|b|^2=1$, $|a|\geq |b|$, $|c|^2+|d|^2=1$, $|c|\geq
|d|$. The overall state of the six ions is
\begin{equation}\label{3}
\begin{array}{lll}
|\psi\rangle_{\rm total} & = &|\phi\rangle_{12}\otimes
|\psi\rangle_{34}\otimes|\psi\rangle_{56}\\
 & = & \alpha ac|eeeeee\rangle_{123456}+\alpha
ad|eeeegg\rangle_{123456}\\
&&+\alpha bc|eeggee\rangle_{123456}+\alpha
bd|eegggg\rangle_{123456}\\
& & +\beta ac|egeeee\rangle_{123456}+\beta
ad|egeegg\rangle_{123456}\\
&&+\beta bc|egggee\rangle_{123456}+\beta
bd|eggggg\rangle_{123456}\\& & +\gamma
ac|geeeee\rangle_{123456}+\gamma
ad|geeegg\rangle_{123456}\\
&&+\gamma bc|geggee\rangle_{123456}+\gamma
bd|gegggg\rangle_{123456}\\& & +\delta
ac|ggeeee\rangle_{123456}+\delta
ad|ggeegg\rangle_{123456}\\
&&+\delta bc|ggggee\rangle_{123456}+\delta
bd|gggggg\rangle_{123456}.\\\end{array}
\end{equation}

One transports adiabatically the ions 3 and  5  to the trap $A$. The purpose of the teleportation protocol is to transmit the arbitrary state $|\phi\rangle$ of the ions 1 and 2, mentioned in Eq.(1), to the ions 4 and 6.

  In
order to make ion 3 to undergo entanglement with ion 1, Alice
performs a C-Not operation between the ion 1 and the ion 3.
Similarly, the same C-Not  operation has been implemented
 between the ion 2 and ion 5. After that  Alice makes
the Hadamard transformations on  ion 1 and ion 2, respectively. So
the state $|\psi\rangle_{\rm total}$ becomes
\begin{equation}\label{4}
\begin{array}{ll}
|\psi'\rangle =&\frac {1}{2}|ee\rangle_{13}[|ee\rangle_{25}(\alpha
ac|ee\rangle_{46}+\beta ad|eg\rangle_{46}\\
 & +\gamma
bc|ge\rangle_{46}+\delta bd|gg\rangle_{46})\\
&+|eg\rangle_{25}(\alpha
ad|eg\rangle_{46}+\beta ac|ee\rangle_{46}\\
& +\gamma bd|gg\rangle_{46}+\delta bc|ge\rangle_{46})\\
&+|ge\rangle_{25}(\alpha
ac|ee\rangle_{46}-\beta ad|eg\rangle_{46}\\
& +\gamma bc|ge\rangle_{46}-\delta bd|gg\rangle_{46})\\
&+|gg\rangle_{25}(\alpha
ad|eg\rangle_{46}-\beta ac|ee\rangle_{46}\\
& +\gamma bd|gg\rangle_{46}-\delta bc|ge\rangle_{46})]\\
&\frac {1}{2}|eg\rangle_{13}[|ee\rangle_{25}(\alpha
bc|ge\rangle_{46}+\beta bd|gg\rangle_{46}\\
 & +\gamma
ac|ee\rangle_{46}+\delta ad|eg\rangle_{46})\\
&+|eg\rangle_{25}(\alpha
bd|gg\rangle_{46}+\beta bc|ge\rangle_{46}\\
& +\gamma ad|eg\rangle_{46}+\delta ac|ee\rangle_{46})\\
&+|ge\rangle_{25}(\alpha
bc|ge\rangle_{46}-\beta bd|gg\rangle_{46}\\
& +\gamma ac|ee\rangle_{46}-\delta ad|eg\rangle_{46})\\
&+|gg\rangle_{25}(\alpha
bd|gg\rangle_{46}-\beta bc|ge\rangle_{46}\\
& +\gamma ad|eg\rangle_{46}-\delta ac|ee\rangle_{46})]\\
&\frac {1}{2}|ge\rangle_{13}[|ee\rangle_{25}(\alpha
ac|ee\rangle_{46}+\beta ad|eg\rangle_{46}\\
 & -\gamma
bc|ge\rangle_{46}-\delta bd|gg\rangle_{46})\\
&+|eg\rangle_{25}(\alpha
ad|eg\rangle_{46}+\beta ac|ee\rangle_{46}\\
& -\gamma bd|gg\rangle_{46}-\delta bc|ge\rangle_{46})\\
&+|ge\rangle_{25}(\alpha
ac|ee\rangle_{46}-\beta ad|eg\rangle_{46}\\
& -\gamma bc|ge\rangle_{46}+\delta bd|gg\rangle_{46})\\
&+|gg\rangle_{25}(\alpha
ad|eg\rangle_{46}-\beta ac|ee\rangle_{46}\\
& -\gamma bd|gg\rangle_{46}+\delta bc|ge\rangle_{46})]\\
&\frac {1}{2}|gg\rangle_{13}[|ee\rangle_{25}(\alpha
bc|ge\rangle_{46}+\beta bd|gg\rangle_{46}\\
 & -\gamma
ac|ee\rangle_{46}-\delta ad|eg\rangle_{46})\\
&+|eg\rangle_{25}(\alpha
bd|gg\rangle_{46}+\beta bc|ge\rangle_{46}\\
& -\gamma ad|eg\rangle_{46}-\delta ac|ee\rangle_{46})\\
&+|ge\rangle_{25}(\alpha
bc|ge\rangle_{46}-\beta bd|gg\rangle_{46}\\
& -\gamma ac|ee\rangle_{46}+\delta ad|eg\rangle_{46})\\
&+|gg\rangle_{25}(\alpha
bd|gg\rangle_{46}-\beta bc|ge\rangle_{46}\\
& -\gamma ad|eg\rangle_{46}+\delta ac|ee\rangle_{46})].\\
\end{array}
\end{equation}

 For the purpose of teleportation,   some measurements on ions
 1 and 3 and ions 2 and 5 have to be made by Alice in trap $A$. After that  Alice
 transmits the measurement outcomes  to Bob, who  controls the trap $B$, via a classical channel. For
 example, if Alice tells Bob that her measurement outcomes are
 $|ee\rangle_{13}$ and $|ee\rangle_{25}$, then Bob can make conclusion that
 the state of ions 4 and  6 is
\begin{equation}
\alpha ac|ee\rangle_{46}+\beta ad|eg\rangle_{46} +\gamma
bc|ge\rangle_{46}+\delta bd|gg\rangle_{46}.
\end{equation}

Assume that the center-of-mass vibration mode in trap $B$ is
initially prepared in the vacuum state $|0\rangle$, so the state of
the system in trap $B$ becomes
\begin{equation}
|\psi(0)\rangle= (\alpha ac|ee\rangle_{46}+\beta ad|eg\rangle_{46}
+\gamma bc|ge\rangle_{46}+\delta bd|gg\rangle_{46})|0\rangle.
\end{equation}

 In the trap B, when a laser  standing wave tuned to the first lower vibrational
sideband was applied to the  ion 6 and the Lamb-Dicke criterion is
satisfied,  the Hamiltonian in an interaction picture reads
\cite{CiracZoller,ZhengSB}
\begin{equation}
H_I=g(c\sigma_+{\rm e}^{-{\rm i}\phi}+c^+\sigma_-{\rm e}^{{\rm
i}\phi}),
\end{equation}
where $c^+$ and $c$ are the creation and annihilation operators of
phonon, $g$ is the Rabi frequency, $\sigma_+=|e\rangle_6\langle g|$
and $\sigma_-=|g\rangle_6\langle e|$, $\phi$ is the phase of this
laser field.  As the Hamiltonian for the quantum system is not a
time-varying,  hence if the laser beam is on for a certain time $t$,
the evolution of the system will be described by the unitary
operator
\begin{equation}
\begin{array}{ll}
U(t)&={\rm e}^{-\mathrm{i}Ht}\\
&=  \cos(gt\sqrt {1+c^+c})|e\rangle_6\langle e|\\
&~~ -{\rm i}\mathrm{e}^{-\mathrm{i}\phi}\frac {\sin(gt\sqrt
{1+c^+c})}{\sqrt
{1+c^+c}}c|e\rangle_6\langle g|\\
&~~-{\rm i}\mathrm{e}^{\mathrm{i}\phi}\frac {\sin(gt\sqrt
{c^+c})}{\sqrt {c^+c}}c^+|g\rangle_6\langle e|\\
&~~+  \cos(gt\sqrt {c^+c})|g\rangle_6\langle g|.\end{array}
\end{equation}
Therefore, after a certain time $t$, we have
\begin{equation}
\begin{array}{l}
|e\rangle_6|0\rangle \rightarrow \cos(gt)|e\rangle_6|0\rangle -{\rm
i e}^{{\rm
i}\phi}\sin(gt)|g\rangle_6|1\rangle,\\
|g\rangle_6|0\rangle \rightarrow |g\rangle_6|0\rangle.
\end{array}
\end{equation}
So, when the laser beam is applied on the ion 6 for the time
interval $t_1$, the
 state $|\psi(0)\rangle$  evolves into
\begin{equation}
\begin{array}{ll}
|\psi(t_1)\rangle=&[\alpha ac \cos(gt_1)|ee\rangle_{46}+\beta ad
|eg\rangle_{46}\\
&+\gamma bc \cos(gt_1)|ge\rangle_{46}+\delta bd
|gg\rangle_{46}]|0\rangle\\
&-{\rm i e}^{{\rm i}\phi}\sin(gt_1)(\alpha ac|eg\rangle_{46}+\gamma
bc|gg\rangle_{46})|1\rangle.\\
\end{array}
\end{equation}

Let us choose  $t_1=\frac {1}{g}\arccos |\frac {d}{c}|$, then  we
have
\begin{equation}
\begin{array}{ll}
|\psi(t_1)\rangle'=&[\mathrm{e}^{\mathrm{i}(\theta_1+\theta_3)}|ad|
\alpha
|ee\rangle_{46}+\mathrm{e}^{\mathrm{i}(\theta_1+\theta_4)}|ad|
\beta|eg\rangle_{46}\\
&+\mathrm{e}^{\mathrm{i}(\theta_2+\theta_3)} |bd|\gamma
|ge\rangle_{46}+ \mathrm{e}^{\mathrm{i}(\theta_2+\theta_4)}|bd|
\delta|gg\rangle_{46}]|0\rangle\\
&-{ \mathrm{i e}}^{{\mathrm{ i}}\phi}\sin(gt_1)(\alpha
ac|eg\rangle_{46}+\gamma
bc|gg\rangle_{46})|1\rangle,\\
\end{array}
\end{equation}
where $a=|a|e^{\rm i\theta_1}$,  $b=|b|e^{\rm i\theta_2}$,
$c=|c|e^{\rm i\theta_3}$, $d=|d|e^{\rm i\theta_4}$. Then  Bob makes  a
measurement on phonon. If the result $|1\rangle$ is obtained, then
the teleportation fails. When the measurement result is $|0\rangle$,
the state of the ions 4 and 6  can be written as
\begin{equation}
\begin{array}{ll}
|\psi\rangle'=&[\mathrm{e}^{\mathrm{i}(\theta_1+\theta_3)}|ad|
\alpha
|ee\rangle_{46}+\mathrm{e}^{\mathrm{i}(\theta_1+\theta_4)}|ad|
\beta|eg\rangle_{46}\\
&+\mathrm{e}^{\mathrm{i}(\theta_2+\theta_3)} |bd|\gamma
|ge\rangle_{46}+ \mathrm{e}^{\mathrm{i}(\theta_2+\theta_4)}|bd|
\delta|gg\rangle_{46}].\\
\end{array}
\end{equation}

After that, Bob applies a laser standing wave tuned to the first lower vibrational
sideband on the ion 4. By the same argument used to ion 6, after an interaction time $t_2$, we have
\begin{equation}
\begin{array}{l}
|e\rangle_4|0\rangle \rightarrow \cos(gt_2)|e\rangle_4|0\rangle
-{\rm i e}^{{\rm i}\phi}\sin(gt
_2)|g\rangle_4|1\rangle,\\
|g\rangle_4|0\rangle \rightarrow |g\rangle_4|0\rangle.
\end{array}
\end{equation}
Hence, the state stated in Eq.(13) becomes
\begin{equation}
\begin{array}{ll}
|\psi(t_2)\rangle=&[\mathrm{e}^{\mathrm{i}(\theta_1+\theta_3)}|ad|
\alpha
\cos(gt_2)|ee\rangle_{46}\\&+\mathrm{e}^{\mathrm{i}(\theta_1+\theta_4)}|ad|
\beta\cos(gt_2)|eg\rangle_{46}\\
&+\mathrm{e}^{\mathrm{i}(\theta_2+\theta_3)} |bd|\gamma
|ge\rangle_{46}
\\&+ \mathrm{e}^{\mathrm{i}(\theta_2+\theta_4)}|bd|
\delta|gg\rangle_{46}]|0\rangle\\
&-{ \mathrm{i e}}^{{\mathrm{
i}}\phi}\sin(gt_2)[\mathrm{e}^{\mathrm{i}(\theta_1+\theta_3)}
|ad|\alpha|ge\rangle_{46}\\
&+\mathrm{e}^{\mathrm{i}(\theta_1+\theta_4)}|ad|\beta|gg\rangle_{46}]|1\rangle.\\
\end{array}
\end{equation}
With the choice $t_2=\frac {1}{g}\arccos |\frac {b}{a}|$, we have
\begin{equation}
\begin{array}{ll}
|\psi(t_2)\rangle'=&\mathrm{e}^{\mathrm{i}(\theta_1+\theta_3)}|bd|[
\alpha
 |ee\rangle_{46}\\&+\mathrm{e}^{\mathrm{i}(\theta_4-\theta_3)}\beta
|eg\rangle_{46}\\
&+\mathrm{e}^{\mathrm{i}(\theta_2-\theta_1)} \gamma |ge\rangle_{46}
\\&+ \mathrm{e}^{\mathrm{i}(\theta_2+\theta_4-\theta_1-\theta_3)}
\delta|gg\rangle_{46}]|0\rangle\\
&-{ \mathrm{i e}}^{{\mathrm{
i}}\phi}\sin(gt_2)[\mathrm{e}^{\mathrm{i}(\theta_1+\theta_3)}
|ad|\alpha|ge\rangle_{46}\\
&+\mathrm{e}^{\mathrm{i}(\theta_1+\theta_4)}|ad|\beta|gg\rangle_{46}]|1\rangle.\\
\end{array}
\end{equation}

Now Bob makes a measurement on phonon again. If the measurement
result is $|1\rangle$, it means that the  teleportation fails. When
the measurement outcome $|0\rangle$ is obtained, the state of the
ions 4 and 6  can be written as
\begin{equation}
\begin{array}{l}\alpha
 |ee\rangle_{46}+\mathrm{e}^{\mathrm{i}(\theta_4-\theta_3)}\beta
|eg\rangle_{46} +\mathrm{e}^{\mathrm{i}(\theta_2-\theta_1)} \gamma
|ge\rangle_{46}
\\+ \mathrm{e}^{\mathrm{i}(\theta_2+\theta_4-\theta_1-\theta_3)}
\delta|gg\rangle_{46}. \end{array}\end{equation}

Under the basis $\{|ee\rangle_{46}, |eg\rangle_{46},
|ge\rangle_{46}, |gg\rangle_{46}\}$, a collective unitary
transformation
\begin{equation}
U=\left (\begin{array}{cccc} 1&0&0&0\\0&e^{-i\phi_1}&0&0\\0&0&e^{-i\phi_2}&0\\0&0&0&e^{-i(\phi_1+\phi_2)}\\
\end{array}\right )
\end{equation}
is made, then the state  in Eq.(17) becomes
\begin{equation}
\alpha
 |ee\rangle_{46}+\beta
|eg\rangle_{46} +\gamma |ge\rangle_{46} + \delta|gg\rangle_{46}.
\end{equation}
Here $\phi_1=\theta_4-\theta_3$, $\phi_2=\theta_2-\theta_1$. The
state in Eq.(19) is just the state which we want to teleport.

Without any difficult one can show that the successful probability
of the teleportation protocol is $4|bd|^2$. When $|a|=|b|$,
$|c|=|d|$, the probability of successful teleportation is 1.

In summary, we propose a scheme for probabilistic teleportation of an
unknown two-particle state of general formation in ion trap. We hope that this teleportation
protocol  of  two-particle state can be realized experimentally
with presently available techniques in the future.

\end{document}